\documentclass{entcs} 
\usepackage{entcsmacro}
\input pdfcolor.tex 
\usepackage[utf8]{inputenc}
\usepackage[english]{babel}

\usepackage{amsmath}
\usepackage{amssymb}
\usepackage{bm}
\usepackage{listings}
\usepackage{color}
\usepackage{url}
\usepackage{subfigure}
\usepackage{multirow}
\usepackage{bussproofs}
\usepackage{algorithmic}
\usepackage{mathtools}
\usepackage{intmacros}
\usepackage{wrapfig}

\usepackage{article}

\newcommand{\Until}{\mathsf{U}}
\newcommand{\Always}{\mathsf{G}}
\newcommand{\Eventually}{\mathsf{F}}

\def\lastname{Ishii Yonezaki Goldsztejn}
\begin{document}
\begin{frontmatter}
\title{Monitoring Bounded LTL Properties\\ Using Interval Analysis} 
  \author{Daisuke Ishii\thanksref{email1}}
  \author{Naoki Yonezaki\thanksref{email2}}
  \address{Tokyo Institute of Technology, Tokyo, Japan}
  \author{Alexandre Goldsztejn\thanksref{email3}}
  \address{CNRS, IRCCyN, Nantes, France}
  \thanks[email1]{Email: \href{mailto:dsksh@acm.org} {\texttt{\normalshape dsksh@acm.org}}}
  \thanks[email2]{Email: \href{mailto:yonezaki@cs.titech.ac.jp} {\texttt{\normalshape yonezaki@cs.titech.ac.jp}}}
  \thanks[email3]{Email: \href{mailto:alexandre.goldsztejn@gmail.com} {\texttt{\normalshape alexandre.goldsztejn@gmail.com}}}
\begin{abstract} 
Verification of temporal logic properties plays a crucial role in proving the desired behaviors of hybrid systems. 
In this paper, we propose an interval method for verifying the properties described by a bounded linear temporal logic.
We relax the problem to allow outputting an inconclusive result when verification process cannot succeed with a prescribed precision, and present an efficient and rigorous monitoring algorithm that demonstrates that the problem is decidable.
This algorithm performs a forward simulation of a hybrid automaton, detects a set of time intervals in which the atomic propositions hold, and validates the property by propagating the time intervals.
A continuous state at a certain time computed in each step is enclosed by an interval vector that is proven to contain a unique solution.
In the experiments, we show that the proposed method provides a useful tool for formal analysis of nonlinear and complex hybrid systems.
\end{abstract}
\end{frontmatter}

\section{Introduction}

Reasoning of the temporal logic properties in a hybrid system is a challenging and important task that lies in the intersection among computer science, numerical analysis, and control theory.
Various methods for falsification of hybrid systems with temporal properties have been developed, e.g., \cite{Plaku2009,Nghiem2010,David2012,Zuliani2013}, and these methods enable verification of various properties (e.g., safety, stability, and robustness) of large and complex systems. The state-of-the-art tools are based on numerical simulations whose numerical errors often produce a qualitatively wrong result and become problematic even in a statistical evaluation.

A fundamental process in formal methods for hybrid systems is computation of rigorously approximated reachable sets. 
The techniques 
based on interval analysis (Section~\ref{s:interval}) have shown practicality in the reachability analysis of nonlinear and complex hybrid systems~\cite{Eggers2008,Collins2008,Ramdani2011,Ishii2011,Chen2012,Gao2013:SMODE,Gao2014}. 
In these frameworks, computation is \emph{$\delta$-complete} \cite{Gao2012}: function values are allowed to be perturbed within a predefined $\delta \in \PosRealSetS$, and, by setting bounds in a problem description, many generically undecidable problems become decidable.
The $\delta$-complete verification of generic properties other than reachability is a challenging topic.

In this paper, we present an interval method for verifying bounded linear temporal logic (BLTL) properties (Section~\ref{s:bltl}) for a class of hybrid automata (Section~\ref{s:ha}).
Our method computes three values in a reliable manner: the algorithm assures the soundness using interval analysis when the result $\Valid$ or $\Unsat$ is output; otherwise, $\Unknown$ is output when the verification process reaches a prescribed precision threshold.
We present an algorithm (Section~\ref{s:method}) based on the forward simulation of a system.
It encloses a trajectory with a set of \emph{boxes} (i.e., interval vectors) and also ensures the unique existence property (i.e., we ensure that a unique state is enclosed in a box corresponding to each initial value) for each step of the simulation.
For each atomic proposition involved in a property $\phi$ to verify, the algorithm obtains an inner and outer approximation of the time intervals in which the proposition holds.
Next, the set of time intervals is modified according to the syntax of the property $\phi$, and finally the algorithm checks whether $\phi$ holds at the initial time.
Using our implementation, we show that nonlinear models are verified and the numerical robustness of a trajectory is assured (Section~\ref{s:ex}).
Although our method is simple, 
it enables reliable analysis of a set of trajectories
and provides a foundation for validated model checking and controller synthesis.

\section{Related Work} \label{s:related}

Many previous studies have applied interval methods to reachability analysis of hybrid systems~\cite{Eggers2008,Collins2008,Ramdani2011,Ishii2011,Chen2012,Gao2013:SMODE,Gao2014}.
The outcome of these methods is an over-approximation of a set of reachable states with a set of boxes. 
%
In interval analysis, a computation often provides a proof of unique existence of a solution within a resulting interval.
This technique also applies in interval-based reachability analysis~\cite{Ishii2011,Goubault2014}, but it is not considered in most of the methods for hybrid systems.
Our method enforces the use of the proof to verify more generic temporal properties.


Reasoning of real-time temporal logic has been a research topic of interest~\cite{Alur1996,Shultz1997}. 
Numerical method for \emph{falsification} of a temporal property is straightforward~\cite{Maler2003}.
It simulates a trajectory of a bounded length and checks the satisfiability of the negation of the property described by a bounded temporal logic.
This paper presents an interval extension of this falsification method.

A tree-search method for searching witness trajectories~\cite{Plaku2009}, a falsification method based on a Monte-Carlo optimization technique~\cite{Nghiem2010}, and statistical model checking methods~\cite{David2012,Zuliani2013} have been proposed.
These methods have been shown their practicality in the verification of realistic nonlinear models;
however, their implementations are based on numerical simulations and might suffer from numerical error.
Applications of our interval method include an integration with these statistical methods to achieve both reliability and practicality.
An integrated statistical and interval method was also proposed in \cite{Wang2014} for reachability analysis.

Notions of \emph{robustness} have been proposed to facilitate the simulation-based verification of temporal properties~\cite{Fainekos2006a,Donze2010,Nghiem2010}.
In these works, the degree of robustness is represented as a distance between a trajectory and a region where a proposition holds.
A non-robust trajectory, which is computed numerically, is likely to be inconsistent with the considered model due to numerical errors.
Our method ensures a robustness rigorously by verifying that a trajectory intersects with each boundary in the state space.

There exist a few methods for model checking of temporal logic properties~\cite{Podelski2006,Cimatti2014}.
\cite{Podelski2006} proposed a method specialized in stability properties, which is described as a specific form of temporal logic formula.
\cite{Cimatti2014} proposed a method that translates a verification problem into a reachability problem with the $k$-Liveness scheme,
which is incomplete in general settings.
Our method can be viewed as a bounded model checking method that validates a bounded temporal property by ensuring that all trajectories that emerge from an initial interval value satisfy the property.


\section{Interval Analysis}
\label{s:interval}

This section introduces selected topics and techniques based on interval analysis~\cite{Moore1966,Neumaier1990}.
The techniques are used in the proposed method in Section~\ref{s:method}.

\subsection{Basic Notions and Techniques} \label{s:ia}

A (bounded) \textit{interval} $\a = [\LB{a}, \UB{a}]$ is a connected set of real numbers $\{b \in \RealSet ~|~ \LB{a} \leq b \leq \UB{a}\}$ and $\IntSet$ denotes the set of intervals.
For an interval $\a$,
$\LB{a}$ and $\UB{a}$ denote the lower and upper bounds;
the width is defined as $\UB{a}-\LB{a}$; and
$\Inter{\a}$ denotes the interior %
$\{b \in \RealSet ~|~ \LB{a} < b < \UB{a}\}$.
%
$[a]$ denotes a point interval $[a,a]$.
For intervals $\a$ and $\b$,
$\Dist{\a}{\b}$ denotes the 
hypermetric between the two, i.e., $\max ( |\UB{a}-\UB{b}| , |\LB{a}-\LB{b}| )$.
%
For a set $S \subset \RealSet$, $\Box S$ denotes the interval $[\inf S, \sup S]$.
All of these definitions are naturally extended to interval vectors;
an $n$-dimensional \textit{box} (or interval vector) $\a$ is a tuple of $n$ intervals
$(\a_1, \ldots, \a_n)$, and
$\IntSet^n$ denotes the set of $n$-dimensional boxes.
For $a\in\RealSet^n$ and $\a\in\IntSet^n$,
we use the notation $a \in \a$, which is interpreted as $\ForAll{i}{\{1,\ldots,n\}} ~ a_i \in \a_i$.
In an actual implementation, the bounds of intervals should be machine-representable floating-point numbers and other real values are rounded in the appropriate directions.


For a function $f : \RealSet^n \to \RealSet$,
$\f : \IntSet^n \to \IntSet$ is known as an \emph{interval extension} of $f$
if and only if it satisfies the containment condition
$\ForAll{\a}{\IntSet^n} ~ \ForAll{a}{\a} ~ ( f(a) \in \f(\a) )$.
This definition is generalized to function vectors $\f : \RealSet^n \to \RealSet^{n_f}$ where $n_f \in \NatSet_{>1}$.
Given intervals $\a,\b\in\IntSet$, interval extensions of four operators $\circ \in \{+,-,\ast,/\}$ can be computed as $\Box\{\LB{a}\circ\LB{b}, \LB{a}\circ\UB{b}, \UB{a}\circ\LB{b}, \UB{a}\circ\UB{b}\}$ (we assume $0 \not\in \b$ for division).

For arbitrary intervals $\a,\b,\d\in\IntSet$, the \emph{extended division} 
$\Box\{d\in\d ~|~ \Exists{a}{\a}~\Exists{b}{\b}~ a = b d\}$
can be implemented as follows (see Section~4.3 of \cite{Neumaier1990}):
\[
	\ExtDiv(\a,\b,\d) := 
	\begin{cases}
		\a/\b \cap \d & \text{if}~ 0 \not\in \b\\
		\Box(\d \setminus (\LB{a}/\LB{b}, \LB{a}/\UB{b})) & \text{if}~ \a > 0 \in \b\\
		\Box(\d \setminus (\UB{a}/\UB{b}, \UB{a}/\LB{b})) & \text{if}~ \a < 0 \in \b\\
		\d & \text{if}~ 0 \in \a,\b
	\end{cases}
\]
In the second and third cases, when $\LB{b}=0$ (resp. $\UB{b}=0$), we set $\LB{a}/\LB{b}$ and $\UB{a}/\LB{b}$ as $-\infty$ and $\infty$ (resp. $\LB{a}/\UB{b}$ and $\UB{a}/\UB{b}$ as $\infty$ and $-\infty$).


Given a differentiable function $f(a) : \RealSet \to \RealSet$ and a domain interval $\a$,
a root $\tilde{a} \in \a$ of $f$ such that $f(\tilde{a}) = 0$ is included in the result of an \emph{interval Newton operator}
\[
	\a \cap \left( \hat{a} - {\f(\hat{a})} / {\tfrac{d}{da}\f(\a)} \right),
\]
where $\hat{a} \in \a$, and $\f$ and $\tfrac{d}{da}\f$ are interval extensions of $f$ and the derivative of $f$.
Iterative applications of the operator will converge.
Let $\a'$ be the result of applying the operator to $\a$. If $\a' \subseteq \Inter{\a}$ holds, a unique root exists in $\a'$.

\subsection{ODE Integration}
\label{s:ode}

An initial value problem (IVP) for an ordinary differential equation (ODE) is specified by a triple $\Struct{t_0, x_0, F}$ consisting of an initial value $x_0 \in \RealSet^n$ at time $t_0 \in \RealSet$ and a flow function $F : \RealSet^n \to \RealSet^n$ (assume Lipschitz continuity).
Given a time interval $\t \in \IntSet$ and a \emph{continuous trajectory} $\tilde{x}(t) : \t \to \RealSet^n$, the satisfaction for IVP-ODEs is defined as
\begin{equation*}
	\tilde{x}, \t \models \Struct{t_0,x_0,F}
	~~\text{iff}~~
	\tilde{x}(t_0) = x_0 \land \ForAll{\tilde{t}}{\t} ~ \tfrac{d}{dt}\tilde{x}(\tilde{t}) = F(\tilde{x}(\tilde{t})).
\end{equation*}
Given $\t_0 \in \IntSet$ and $\x_0 \in \IntSet^n$, we can consider a parametric IVP-ODE $\Struct{\t_0, \x_0, F}$, where the initial condition is parameterized, and its satisfaction relation is defined as
\begin{equation*}
	\tilde{x}, \t \models \Struct{\t_0,\x_0,F} ~~\text{iff}~~
	\Exists{t_0}{\t_0} ~ \Exists{x_0}{\x_0} ~ \tilde{x}, \t \models \Struct{t_0,x_0,F}.
\end{equation*}
$\Trajs_\mathbm{t}\Struct{\mathbm{t}_0, \mathbm{x}_0, F}$ denotes the set of satisfied trajectories on $\t$.

Using the tools based on the interval Taylor methods, e.g.,  CAPD\footnote{\url{http://capd.ii.uj.edu.pl/}} and VNODE~\cite{Nedialkov2006}, we can obtain an interval extension $\XC : \IntSet\to\IntSet^n$ of solution trajectories in $\Trajs_\mathbm{t}{\Struct{\mathbm{t}_0, \mathbm{x}_0, F}}$.
Given $\t' \in {\IntSet}$, such tools compute a value $\XC(\t')$ by performing the stepwise integration of the flow function $F$ from the initial time $\t_0$ to time $\UB{t}'$. 
%
In the stepwise computation of the interval Taylor methods, the \emph{unique existence} of a solution is verified for a box enclosure computed in each step based on the Picard-Lindel\"{o}f operator and Banach's fixpoint theorem.
%
Accordingly, when an interval enclosure $\XC(\t')$ (assume $\LB{t}' \geq \UB{t}_0$) is computed with an interval Taylor method, the following property holds:
\begin{equation*} \label{e:ode:unique}
	\ForAll{t_0}{\t_0} ~ \ForAll{x_0}{\x_0} ~ 
	\Exists{\text{unique\ }\tilde{x}}{(\t' \to \XC(\t'))} ~ \tilde{x}, \t' \models \Struct{t_0, x_0, F}.
\end{equation*}

In principle, if $F$ is Lipschitz continuous and we can assume an arbitrary precision, we obtain an arbitrary narrow interval enclosure $\XC([t])$ for $t \in \RealSet$.
However, since the implementations use machine-representable real numbers, it may fail to compute an enclosure in the process that verifies the unique existence property, even with the smallest step size.

\section{Hybrid Automata}
\label{s:ha}

We model a hybrid system as a hybrid automaton~\cite{Alur1995}.
For simplicity in this paper, we consider deterministic systems, i.e., the location invariant is the negation of guard conditions and two guards do not overlap in a location.
The proposed method can be extended to handle non-deterministic systems, e.g., by enumerating possible paths and computing a trajectory enclosure for each path.

\begin{definition}
A \emph{hybrid automaton} is a septet
\begin{equation*}
	\HA :=
	\bigl( \LocSet, x, X, \Init, 
	\{\Flow_q\}_{q\in\LocSet},
	\{\Grd_{q,q'}\}_{q\in\LocSet,q'\in\LocSet},
	\{\Rst_{q,q'}\}_{q\in\LocSet,q'\in\LocSet}
	\bigr),
\end{equation*}
that consists of the following components:
\begin{itemize}
\setlength{\parskip}{0pt}
\setlength{\itemsep}{0pt}
\item A finite set of \emph{locations} $\LocSet = \{q_1,\ldots,q_{n_q}\}$.
\item A vector of real-valued variables $x = (x_1,\ldots,x_n)$.
\item A domain $X \subseteq \IntSet^n$ for the valuation of the variables.
\item A set of initial values $\Init \subseteq \{q\}\Times X$ where $q \in \LocSet$.
\item A set of \emph{vector fields} $\Flow_q : X \to X$ (assume Lipschitz continuity).
\item A set of \emph{guards} $\Grd_{q,q'} \subseteq X$ described by a condition of the form $g(x) = 0 \LAnd h(x) < 0$ where $g,h : X \to \RealSet$.
\item A set of \emph{reset functions} $\Rst_{q,q'} : X \to X$.
\end{itemize}
\end{definition}

Behaviors of the states $\sigma \in \LocSet\Times X$ over the timeline are formalized as \emph{trajectories}.
In this work, we assume that
there are no consecutive multiple discrete changes.
\begin{definition}
Given an $\HA$, an initial state $\Struct{q_0,s_0} \in \Init$, and a time interval $\t = [0,t_\Max]$ ($t_\Max \in \PosRealSet$),
a state at each time $t \in \t$ is determined as a \emph{trajectory} 
$\Struct{\LTraj,\CTraj}$, which consists of 
a \emph{location trajectory} $\LTraj : \t \to \LocSet$ and
a \emph{continuous state trajectory} $\CTraj : \t \to X$.
The value of the trajectory is defined recursively as follows:
\begin{align*}
	\Struct{\LTraj(0), \CTraj(0)} :=& ~\Struct{q_0, s_0},\\ 
	\Struct{\LTraj(t), \CTraj(t)} :=& ~\sigma \in \LocSet\Times X
	~~\text{s.t. $\Exists{t'}{(0,t)} ~\Exists{\sigma'}{\LocSet\Times X} ~ \Struct{\LTraj(t'),\CTraj(t')} \xrightarrow{t-t'} \sigma' \xrightarrow{0} \sigma$}, \span
\end{align*}
where the relation $\sigma_1\xrightarrow{t}\sigma_2 \in (\LocSet\Times X) \Times \t \Times (\LocSet\Times X)$ is given by the following rules:
\begin{equation*}
	\frac{\Grd_{q,q'}(s) \quad \Rst_{q,q'}(s) = s' 
	}
  { \Struct{q, s} \xrightarrow{0} \Struct{q', s'} }
  \qquad
  \frac{
    \begin{array}{l}
      f(0) = s \quad
	  \ForAll{\tilde{t}}{[0, t]} ~
	  \frac{d}{dt}f(\tilde{t}) = \Flow_q(f(\tilde{t})) \\[-.5em]
	  \hspace{4.6em}
	  \ForAll{\tilde{t}}{[0, t)} ~
	  \ForAll{q'}{\LocSet} ~ \neg G_{q,q'}(f(\tilde{t}))
	\vspace*{-.3em}
    \end{array}
  }
  {\Struct{q, s} \xrightarrow{t} \Struct{q, f(t)}}
\end{equation*}
Note that the second rule also applies for $t=0$.
The set of trajectories of length $t_\Max$ is denoted by $\Trajs_{t_\Max}(\HA)$.
\end{definition}

When a discrete change $\xrightarrow{0}$ is applied at time $t$, the state $\Struct{\LTraj(t),\CTraj(t)}$ overwrites the state $\sigma'$ before the discrete change.
%
An $\HA$ has a unique trajectory $\Struct{\LTraj,\CTraj}$ starting from an initial state $\Struct{q_0, s_0} \in \Init$ because we have assumed that two guards do not hold simultaneously.
%

\begin{figure*}[t]
\includegraphics[width=\linewidth]{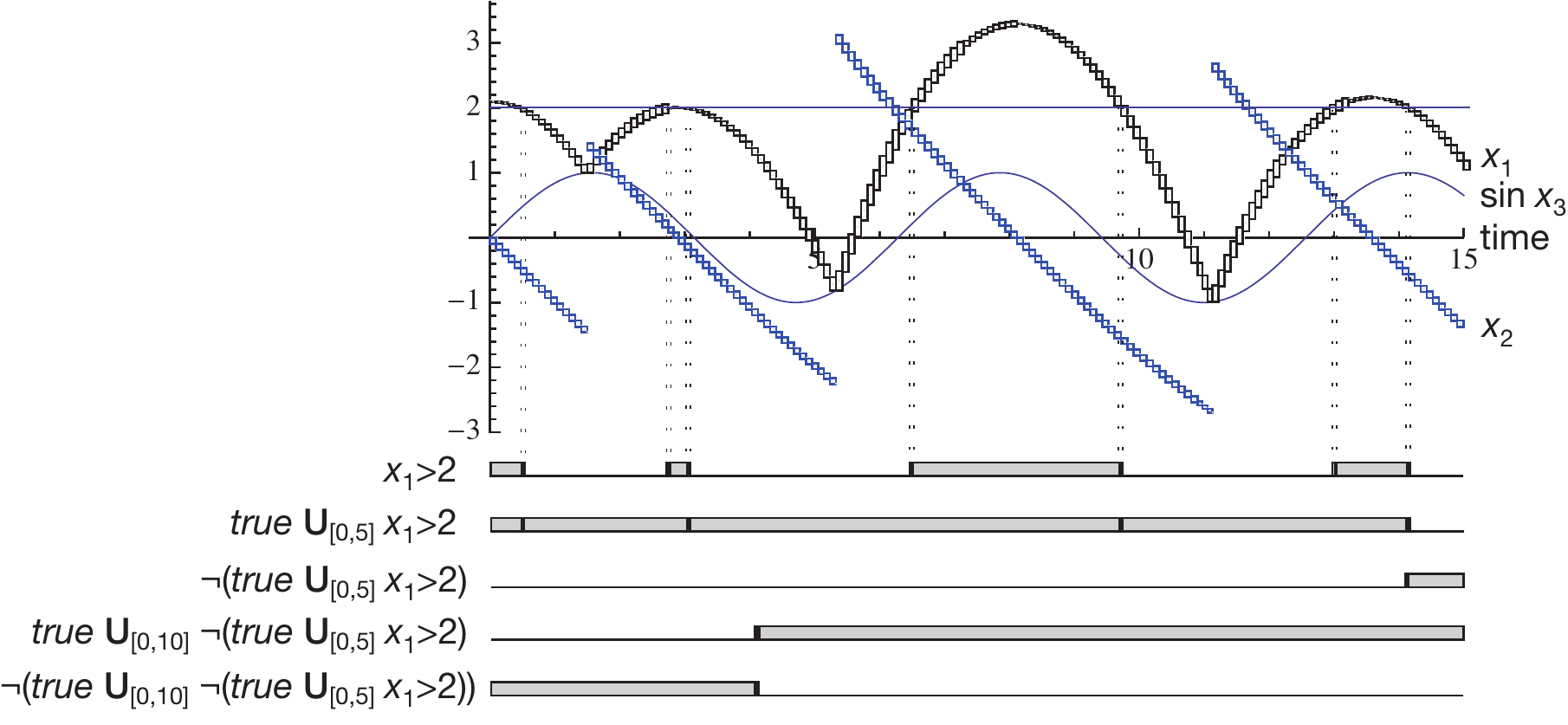}
\caption{Verification of the bouncing ball example}
\label{f:bb}
\end{figure*}

\begin{example} \label{ex:bb}
We model a bouncing ball on a moving table as an $\HA$:
\begin{align*}
	x &:= (x_1,x_2,x_3) \in X := ([-1,10],[-10,10],[0,1000]) \span\span\\
	L &:= \{q\} \span\span\\
	\Init &:= \{q\} \Times ([2,7],[0],[0]) \span\span\\
	F_q &:= (x_2, -1 + 0.04 x_2^2 \,\mathrm{sgn}\, x_2, 1) \span\span \\
	G_{q,q} &:= x_1 - \sin x_3 = 0 \,\land\, x_2-\cos x_3 < 0 \span\span\\
	R_{q,q} &:= (x_1, -0.9 x_2 + 1.9 \cos x_3, x_3) \span\span
\end{align*}
Variables $x_1,x_2,$ and $x_3$ represent the height and velocity of the ball, and the (global) time, respectively.
Air resistance is considered in the dynamics.
The height of the table sinusoidally oscillates within $[-1,1]$ and is represented as $\sin x_3$.
The second proposition of the guard is to forbid the guard to hold right after a discrete change.
Possible trajectories of $x_1$ and $x_2$ are illustrated in Figure~\ref{f:bb}.
%
%
\end{example}

\section{Bounded Linear Temporal Logic}
\label{s:bltl}

We consider a fragment~\cite{Maler2003} of the real-time metric temporal logic~\cite{Alur1996} such that the temporal modalities are bounded by an interval $\t = [\LB{t},\UB{t}]$ such that the bounds $\LB{t},\UB{t}$ are in $\RatSet$.
We refer to the logic \emph{bounded linear temporal logic} (BLTL) as in \cite{Zuliani2013}.

\begin{definition}
We consider constraints in the real domain as atomic propositions.
The syntax of the BLTL formulae is defined by the grammar
\begin{equation*}
	\phi ::= \True ~|~ \Prop ~|~ \phi \lor \phi ~|~ \neg \phi ~|~ \phi\,\Until_{\ttt}\,\phi
	\qquad
	p ::= f(x) < 0 ~|~ f(x) \leq 0 
\end{equation*}
where $\Prop$ belongs to a set of \emph{atomic propositions} $\APSet_\phi$, $\Until_{\ttt}$ is the ``until'' operator bounded with a non-empty positive time interval $\t \in {\IntSet}$, 
$x = (x_1,\ldots,x_n)$ is a vector of variables, and $f : \RealSet^{n} \to \RealSet$.
We use the standard abbreviations, e.g., $\phi_1\land\phi_2 := \neg(\neg\phi_1\lor\neg\phi_2)$, $\Eventually_\ttt \phi := \True\,\Until_\ttt\,\phi$ (``eventually''), and $\Always_\ttt\phi := \neg\Eventually_\ttt\neg\phi$ (``always'').
An equation $f(x) = 0$ can be encoded as $f(x) \leq 0 \land -f(x) \leq 0$.
\end{definition}

\subsection{Semantics}

The necessary length $\Norm{\phi}$ of trajectories for checking a formula $\phi$ is inductively defined by the structure of the formula:
\begin{align*}
	\Norm{p} &:= 0 & 
	\Norm{\phi_1 \lor \phi_2} &:= \Max\,(\Norm{\phi_1},\Norm{\phi_2}) \\
	\Norm{\neg \phi} &:= \Norm{\phi} &
	\Norm{\phi_1 \,\Until_{\ttt}\, \phi_2} &:= \Max\,(\Norm{\phi_1},\Norm{\phi_2}) + \UB{t}
\end{align*}
A map $\Obs : \APSet_\phi \to \PS{X}$ corresponds each proposition $\Prop \in \APSet_\phi$ to a set $\Obs(\Prop) = \{s \!\in\! X ~|~ p(s)\}$.
Let $\Struct{\LTraj,\CTraj}$ be a trajectory in $\Trajs_{\Norm{\phi}}(\HA)$ and $\phi$ be a BLTL property.
We have a satisfaction relation defined as follows:
\begin{align*}
	\CTraj,t &\models \True &&\\
	\CTraj,t &\models \Prop && \text{iff}~~ \CTraj(t) \in \Obs(\Prop)\\
	\CTraj,t &\models \phi_1 \lor \phi_2 && \text{iff}~~ \CTraj,t \models \phi_1 \LOr \CTraj,t \models \phi_2\\
	\CTraj,t &\models \neg \phi && \text{iff}~~ \CTraj,t \not\models \phi\\
	\CTraj,t &\models \phi_1 \,\Until_\ttt\, \phi_2 
	&& \text{iff}~~ \Exists{t'}{(t+\t)} ~ \CTraj,t' \models \phi_2 \LAnd 
	(\ForAll{t''}{[t,t']} ~ \CTraj,t'' \models \phi_1)
\end{align*}
$\phi_1\,\Until_\ttt\,\phi_2$ intuitively means that (assuming we are at time $t$) $\phi_2$ will hold within the time interval $t\!+\!\t$ and $\phi_1$ always hold until then.
We also have a validation relation defined as:
\[
	\HA \models \phi ~~\text{iff}~~ \ForAll{\Struct{\LTraj,\CTraj}}{\Trajs_{\Norm{\phi}}(\HA)} ~ \CTraj,0 \models \phi
\]

\subsection{Method for Monitoring BLTL Formulae}
\label{s:bltl:monitoring}

Our interval method is based on the method proposed in \cite{Maler2003} that decides whether a trajectory satisfies a BLTL property.
In this section, we explain this basic method.
First, we introduce the notion of consistent time intervals against BLTL formulae.
\begin{definition} \label{th:consistent}
	Let $\Struct{\LTraj,\CTraj}$ be a trajectory of length $t_\Max$ and $\phi$ be a BLTL formula.
	We say that a left-closed and right-open interval $[\LB{t},\UB{t}) \subseteq \PosRealSet$ is \emph{consistent} with $\phi$ iff $\ForAll{t}{[\LB{t},\UB{t})}~ \CTraj,t \models \phi$.
\end{definition}

The satisfiability of a property $\phi$ by a trajectory is checked as follows:
\begin{enumerate}
	\item For each atomic proposition $p$ in $\phi$, monitor the trajectory of length $\Norm{\phi}$ and identify a non-overlapping set of consistent time intervals $T_p = \{\t_1,\ldots,\t_{n_p}\}$.
\item Following the parse tree of $\phi$ in a bottom-up fashion, compute a set of consistent time intervals of $\phi$. For each construct of BLTL, compute as follows:
\begin{align*}
	T_{\neg \phi} &:= 
	\PosRealSet \setminus T_\phi \qquad \qquad \qquad
	T_{\phi_1\lor \phi_2} := T_{\phi_1} \cup T_{\phi_2} \\
	T_{\phi_1 \Until_{\ttt} \phi_2} &:= 
	\{\AlgName{Shift}_{\ttt}(\t_1 \cap \t_2) \cap \t_1 ~|~
	\t_1 \in T_{\phi_1}, \t_2 \in T_{\phi_2} \} 
\end{align*}
where $\AlgName{Shift}_{\ttt}(\s) := [\LB{s}-\UB{t}, \UB{s}-\LB{t}) \cap \PosRealSet$.
\item Check whether the smallest time interval in $T_\phi$ contains time 0.
	If yes, $\phi$ is satisfied; otherwise, it is not satisfied.
\end{enumerate}

\begin{example} \label{ex:bb:bltl}
	We verify the property 
	\[
		\Always_{[0,10]} \Eventually_{[0,5]}\, 2-x_1 \!<\! 0 ~\equiv~ \neg( \True \,\Until_{[0,10]}\, \neg ( \True \,\Until_{[0,5]}\, 2-x_1 \!<\! 0 ))
	\]
	for the model in Example~\ref{ex:bb}.
Computation with the monitoring method (which is extended to an interval method) is illustrated in Figure~\ref{f:bb}.
\end{example}



\section{Interval-Based Simulation and Monitoring Method}
\label{s:method}

In this section, we propose 
an interval extension of the monitoring method in Section~\ref{s:bltl:monitoring}.

In Step (i) of the method,
given an $\HA$ and a BLTL property $\phi$,
we first simulate the $\HA$ for length $\Norm{\phi}$.
Our simulation method computes an over-approximation of a trajectory of $\HA$ in which the existence of a unique trajectory is verified, i.e., 
it verifies the property
\begin{equation*}
\ForAll{\Struct{q_0,x_0}}{\Init}~ \exists \text{unique}\ \Struct{\LTraj,\CTraj} \in \Trajs_{\Norm{\phi}}(\HA)~~
\LTraj(0) = q_0 \LAnd \CTraj(0) = x_0
\end{equation*}
meaning that for each initial value, there exists a unique trajectory of length $\Norm{\phi}$.

Next, our method identifies the time intervals that are consistent with an atomic proposition in $\phi$ (Definition~\ref{th:consistent}).
A consistent time interval $[\LB{t},\UB{t})$ is, in general, not representable in an actual implementation; therefore, we approximate it by a pair of intervals $\u$ and $\u'$ such that each of them encloses the boundaries $\LB{t}$ or $\UB{t}$.
Given an atomic proposition $f(x) \circ 0$ ($\circ \in \{<, \leq\}$) and a trajectory $\CTraj$, we search for a boundary enclosure $\u$ such that $\f(\CTraj(\u)) \ni 0$ where $\f$ is an interval extension of $f$.

In Step (ii), the set of consistent time intervals is updated to be consistent with $\phi$.
This computation requires $\u$ to contain a unique boundary point, and thus a naive over-approximation $\tilde{\u}$ does not suffice because an interval extension $\f(\tilde{\u})$ may contain 0 although $f(\tilde{\u})$ does not contain a boundary or contains several boundaries.
Thanks to interval techniques, our method verifies the unique existence of a boundary in $\u$.
Finally, as an extension of the above property, our method verifies
\begin{equation*}
\ForAll{\Struct{q_0,x_0}}{\Init}~ \exists \text{unique}\ \Struct{\LTraj,\CTraj} \in \Trajs_{\Norm{\phi}}(\HA)~~
\LTraj(0) = q_0 \LAnd \CTraj(0) = x_0 \LAnd
\CTraj,0 \models \phi.
\end{equation*}

The proposed method has some limitations.
First, it is a semi-decision procedure that may output an inconclusive result ($\Unknown$) because of a failure in the verification of unique existence;
both the procedures for enclosing a continuous trajectory and enclosing a time where a discrete change occurs may cause errors.
However, this mechanism is valuable in terms of reliability and complexity of the problem; a non-robust trajectory and a zeno $\HA$ will be rejected as an error in the verification process.
In practice, when addressing a nonlinear $\HA$s, 
the method may only work successfully with a sufficiently small subset $\Init' \subset \Init$ of initial values.
In this way, the method can be still used for $\Sat$/$\Invalid$ checking.
Second, the method is a bounded model-checking method in the sense that the domain $X$ of the variables is bounded, and it assumes a bounded length and a number of discrete changes in a trajectory.

\newpage

\subsection{Main Algorithm}
\label{s:method:main}

\begin{figure}[tb]
\begin{algorithmic}[1]
  \REQUIRE $\HA$, $\phi$
  \ENSURE $\Valid$, $\Unsat$, or $\Unknown$
  \STATE $\t \Asn 0$;\quad
  $q \Asn q_0$;\quad
  $\x \Asn \x_0$;\quad
  $\T \Asn \{\emptyset,\ldots,\emptyset\}$
  %

  \vspace{.5em}
  \WHILETRY{$\LB{t} < \Norm{\phi}$}

    \STATE $\t_c \Asn \Norm{\phi}$


	\FORC{$q' \in \LocSet$} {Find zeros for each edge}
	\STATE $\t'_c \Asn \SearchZero(\XC,F_q,\Grd_{q,q'}, [\LB{t},\Norm{\phi}])$
	\STATE \textbf{if} $\t'_c\neq\emptyset \LAnd \UB{t}'_c < \LB{t}_c$ \textbf{then}
	\STATE \quad $q'' \Asn q'$; \ $\t_c \Asn \t'_c$
	\STATE \textbf{else if} $\UB{t}'_c \geq \LB{t_c}$ \textbf{then} 
	\STATE \quad \textbf{return} $\Unknown$
	\STATE \textbf{end if} 
	\ENDFOR

	\vspace{.5em}
	\FORC{$p = f \circ 0 \in \APSet_\phi$}{Find boundaries of APs}
	\STATE $\t'_c \Asn [\LB{t},\UB{t}_c]$
	\LOOP
	\STATE $\t'_c \Asn \SearchZero(\XC, F_q, f=0, \t'_c)$
	\STATE \textbf{if} $\t'_c = \emptyset$ \textbf{then} \textbf{break} \textbf{end if}
	\STATE $\t'_c \Asn [\UB{t}'_c, \UB{t}_c]$; \quad
	$\T_p := \T_p \cup \{\t'_c\}$ 
    \ENDLOOP
    \ENDFOR

	\vspace{.5em}
	\STATE $\x \Asn \Jump(\XC, \t_c, \Rst_{q,q''}, \APSet_{\phi}, \T)$
	\hfill \COMMENT{Discrete change}
	\STATE $q \Asn q''$; \quad $\t \Asn \t_c$
	\vspace{.5em}

	\STATE \hspace{-1.3em} \textbf{catch error then return} $\Unknown$ \textbf{end try}
  \ENDWHILE

  \vspace{.5em}
  \RETURN $\AlgName{AnalIntervals}(\T)$
\end{algorithmic}
\caption{$\ValidTraj$ algorithm}
\label{a:main}
\end{figure}

Given an $\HA$ and a BLTL property $\phi$, the proposed $\ValidTraj$ algorithm (Figure~\ref{a:main}) outputs the following results:
$\Valid$ that implies $\HA \models \phi$;
$\Unsat$ that implies $\HA \models \neg\phi$; or
$\Unknown$ when the computation is inconclusive.

%
An iteration of the outmost loop corresponds to a continuous phase of the trajectory and a discrete change.
%
%
At Lines~4--11, each guard for a possible transition is evaluated.
The $\SearchZero$ algorithm is described in Section~\ref{s:method:zero} and will return a time interval $\t'_c$ within which the guard is satisfied or $\emptyset$ if no state satisfies the guard.
Next, the algorithm attempts to decide the \emph{earliest} time interval by checking whether $\t'_c$ is strongly less than $\t_c$.
If two guard crossings are too close, so two crossing time intervals overlap, the algorithm returns $\Unknown$.
At Lines~12--19, for each atomic proposition of $\phi$ of the form $f \circ 0$ ($\circ \in \{<,\leq\}$), the algorithm searches for a boundary where the sign of $f$ changes.
Because several boundaries can exist in a continuous phase, the inner loop searches for all of them.
The detected time intervals are saved in the set $\T$ associated with the atomic propositions.
At Lines~20--21, the discrete change between the locations $q$ and $q''$ is computed by evaluating an interval extension of $\Rst_{q,q''}(\XC(\t_c))$.
A jump of state might switch the state of an atomic proposition; if such a switch exists, the $\Jump$ procedure should detect and record it in $\T$.
Finally, boundary points (which are enclosed by intervals) of the consistent time intervals saved in $\T$ are analyzed by the $\AlgName{AnalIntervals}$ procedure (Line~24, Section~\ref{s:method:bltl}).
%
The procedures $\XC$ (Section~\ref{s:ode}) and $\SearchZero$ (Section~\ref{s:method:zero}) may results in errors. These errors are caught by the catch clause at Line~22.

\subsection{Evaluation of BLTL Properties}
\label{s:method:bltl}

The BLTL evaluation explained in Section~\ref{s:bltl:monitoring} can be implemented as a rigorously approximated procedure $\AlgName{AnalIntervals}$.

First, we approximate a set of consistent time intervals $T_\phi = \{\t_1,\ldots, \t_{n_\phi}\}$ by $\T_\phi = \{\u_1,\u_1', \ldots, \AB \u_{n_\phi},\u_{n_\phi}'\}$ such that $\u_i,\u_i'\in\IntSet$, $\LB{t}_i \in \u_i$, $\UB{t}_i \in \u_i'$, $\UB{u}_i \leq \LB{u}_i'$, and $\UB{u}_i' \leq \LB{u}_{i+1}$, for $i \in \{1,\ldots,n_\phi\}$.
$T=\emptyset$ and $T=\PosRealSet$ are approximated as $\T=\emptyset$ and $\T=\{[0]\}$, respectively.
For $\u_i$, $\u_i'$ in $\T_\phi$ and a continuous trajectory $\CTraj$, $\ForAll{t}{[\UB{u}_i,\LB{u}_i')}~\CTraj,t\models\phi$ holds.

Next, the evaluation on the set of time intervals in Step~(ii) is extended to address the approximated sets.
The set of the inverted time intervals for $\neg \phi$ can be represented as 
$\T_{\neg\phi} = \{[0],\u_1, \ldots, \u_{n_\phi}'\}$ if $\LB{u}_1 > 0$, and
$\T_{\neg\phi} = \{\u_1', \ldots, \u_{n_\phi}'\}$ if $\u_1 = [0]$;
otherwise, the evaluation results in $\Unknown$ if $\u_1 \ni 0$.
The union of two sets of time intervals for $\phi_1\lor\phi_2$ can be implemented as a merge and sort process of the two approximated sets.
The $\AlgName{Shift}_\ttt$ procedure for $\phi_1\Until_{\ttt}\phi_2$ can be implemented as translations of the time intervals $\u_i$ and $\u_i'$ for $\UB{t}$ and $\LB{t}$, respectively.
Some more case analyses should be applied, e.g., when the intervals in $\T_\phi$ become redundant or when they overlap.

Finally, we obtain $\T_\phi$ and conclude that $\phi$ is $\Valid$ if $\UB{u}_1 \leq 0 \leq \LB{u}_1'$; it is $\Unsat$ if $\T_\phi = \emptyset$ or $0 < \LB{u}_1$; or the satisfaction is $\Unknown$ if $0 \in [\LB{u}_1,\UB{u}_1)$.


\subsection{Computation of a Continuous Trajectory}
\label{s:method:cont}

If the system is in a location $q \in \LocSet$ and the value of the state variables is $\x \in \IntSet^n$ at time $\t \in {\IntSet}$,
then the subsequent continuous evolution is specified by an IVP-ODE $\Struct{\t,\x,F_q}$.
Next, we can obtain an interval extension $\XC : {\IntSet} \to \IntSet^n$ of the continuous trajectories in $\Trajs_{[\LB{t},\Norm{\phi}]}{\Struct{\mathbm{t},\mathbm{x},F_q}}$ as described in Section~\ref{s:ode}.

\subsection{Evaluation of Boundary Conditions}
\label{s:method:zero}

\begin{figure}[tb]
\begin{algorithmic}[1]
  \REQUIRE $\XC : \IntSet\to\IntSet^n$, ~
     $F : X \to X$, ~
	 $g\!=\!0 \land h\!<\!0$, ~
	 $\t_\mathrm{init} \in \IntSet$
	 \ENSURE $\t\in\IntSet\cup\{\emptyset\}$
  \PARAM 
  $\epsilon \in \SPosRatSet$, $\theta \in (0,1)$
  \STATE $\t \Asn \t_\mathrm{init}$ 
  \REPEATC{Lower bound reduction}
    \STATE $\t_\mathrm{old} \Asn \t$
	\STATE $\d_g \Asn \Dt(g, \XC, F, \t)$;~ $\d_h \Asn \Dt(h, \XC, F, \t)$
	\STATE $\t \Asn \LB{t}+\ExtDiv(-\g(\XC(\LB{t})),\ \d_g,\ \t-\LB{t})$
	\STATE $\t \Asn \LB{t}+\ExtDiv(-\h(\XC(\LB{t})) - [0,\infty],\ \d_h, \t-\LB{t})$
  \UNTIL{$d(\t_\mathrm{old},\t) \leq \epsilon$}
  \STATE \textbf{if} $\t = \emptyset$ \textbf{then return} $\emptyset$ \textbf{end if}
  \vspace{.5em}
  \STATE $\t \Asn \LB{t}$
  \LOOPC{Unique solution existence verification}
	\STATE $\d_g \Asn \Dt(g, \XC, F, \t)$
    \STATE \textbf{if} $\d_g \ni 0$ \textbf{then error end if}
    \STATE $\t' \Asn \LB{t}-{\g(\XC(\LB{t}))}/{\d_g}$
	  \STATE \textbf{if} $\t' \subseteq \Inter{\t}$
	  \textbf{then} $\t \Asn \t'$; \textbf{break end if}
    \STATE $\t_\mathrm{bak} \Asn \t$
	\STATE $\t \Asn \t_\mathrm{init} \cap \AlgName{Inflate}(\t',1\!+\!\theta)$
	\STATE \textbf{if} $d(\t,\t') > (1\!-\!\theta)\, d(\t',\t_\mathrm{bak})$ \textbf{then error end if}
  \ENDLOOP
  %
  %
  \vspace{.5em}
  \STATE \textbf{if} $\sup \h(\XC(\t)) \geq 0$ \textbf{then error end if}
  \RETURN $\t$
\end{algorithmic}
\caption{\label{a:searchzero} $\SearchZero$ algorithm}
\end{figure}

Guards and atomic propositions can be treated as \emph{boundary conditions}
in the state space $X \subseteq \RealSet^n$ of the form 
\[
	B(x) \ :=\ g(x) = 0 \LAnd h(x) < 0,
\]
where $g : X \to \RealSet$ and $h : X \to \RealSet$.
We propose the $\SearchZero$ algorithm shown in Figure~\ref{a:searchzero} for searching the intersection between a trajectory and a boundary condition.
Inputs to the algorithm consist of an interval extension of the continuous trajectory $\XC : \IntSet\to\IntSet^n$, a vector field of the current location $F : X \to X$, the boundary condition $B(x)$, and a time interval $\t_\mathrm{init} \in\IntSet$ to be searched.
$\SearchZero$ searches for the earliest time interval $\t \subseteq \t_\mathrm{init}$ such that the state $\XC(\t)$ encloses a unique solution of the boundary condition, i.e.,
\begin{equation} \label{e:zero:earliest}
	\t = \Box \bigl\{ \Min \{t \!\in\! \t_\mathrm{init} ~|~ B(\CTraj(t))\} ~|~ \CTraj \!\in\! \Trajs_{\mathbm{t}_\mathrm{init}}{\Struct{\mathbm{t}_0, \mathbm{x}_0, F_q}} \bigr\},
\end{equation}
where $\Struct{\mathbm{t}_0, \mathbm{x}_0, F_q}$ denotes the IVP-ODE of the current location.
Moreover, $\SearchZero$ verifies the following property:
\begin{equation} \label{e:zero:unique}
	\ForAll{\CTraj}{\Trajs_{\mathbm{t}_\mathrm{init}}{\Struct{\mathbm{t}_0, \mathbm{x}_0, F_q}}}~ \exists \text{unique}~ t\!\in\!\t~ B(\CTraj(t))
\end{equation}
Otherwise, $\SearchZero$ returns $\emptyset$ when the boundary condition is unsatisfiable, i.e.,
\begin{equation} \label{e:zero:unsat}
	\ForAll{\CTraj}{\Trajs_{\mathbm{t}_\mathrm{init}}{\Struct{\mathbm{t}_0, \mathbm{x}_0, F_q}}}~ \ForAll{t}{\t_\mathrm{init}}~ \neg B(\CTraj(t)).
\end{equation}


\begin{lemma}
	If $\SearchZero$ returns a non-empty interval $\t$, the properties \eqref{e:zero:earliest} and \eqref{e:zero:unique} hold.
	If it returns $\emptyset$, the property \eqref{e:zero:unsat} holds.
\end{lemma}


To justify the soundness, we describe some details of the algorithm. 
At Lines~2--7, the time interval $\t$ is filtered repeatedly using an interval Newton operator.
At Line~4 (and at Line~11), given a function $g$, the $\Dt$ procedure computes an interval enclosure of the derivative $\frac{d}{dt}g(\CTraj(t))$ over the time interval $\t$ using the chain rule
\[
	\tfrac{d}{dt}g(\CTraj(\t)) = \tfrac{d}{dx}g(\CTraj(\t)) \cdot \tfrac{d}{dt}\CTraj(\t)
	\ \subseteq\ \tfrac{d}{dx}\g(\XC(\t)) \cdot \F(\XC(\t)).
\]
Next, at Lines~5 and 6, the interval Newton is applied.
The extended division (Section~\ref{s:interval}) is used to implement the interval Newton to handle the numerator intervals $\d_g,\d_h$ containing zero.
Because we expand the interval Newton on the lower bound $\LB{t}$ and the extended division encloses the values in the domain $\t-\LB{t}$, the resulting $\t$ is filtered its inconsistent portion, without losing the solutions or being expanded.
%
When the interval Newton results in $\emptyset$, $\SearchZero$ also returns $\emptyset$ to signal the unsatisfiability.
At Line~9, because $\t$ may contain several solutions, $\t$ is reset to the lower bound as a starting value to compute an enclosure of the earliest solution.
At Lines~10--18, $\SearchZero$ applies the interval Newton method with the inclusion test to prove the unique existence of a solution within the contracted interval $\t'$.
This interval Newton verification is repeated with an inflation process of the time interval (see \cite{Goldztejn2010:RC} for a detailed implementation).
When reaching Line~19 with no error, the time interval $\t$ is a sharp enclosure of the first zero of $g(\CTraj(t))=0$. It remains to check that the inequality constraint $h(\CTraj(t))<0$ is satisfied inside $\t$.

When $\SearchZero$ is implemented with machine representable real numbers, or when there is a tangency between the trajectory and the guard constraint, a computation may result in an error.
%
%
Line~12 of $\SearchZero$ may give rise to error if the derivative on an (inflated) time interval contains zero.
At Line~17, we limit the number of iterations according to whether the inflation ratio reaches the threshold as proposed in \cite{Goldztejn2010:RC}.


\section{Experiments}
\label{s:ex}

We have implemented the proposed method and experimented on two examples to confirm the effectiveness of the method.
Experiments were run using a 2.4GHz Intel Core i5 processor with 16GB of RAM.

\subsection{Implementation}
\label{s:impl}

We have implemented the proposed algorithms in Figures~\ref{a:main} and \ref{a:searchzero} in OCaml and C/C++.
The CAPD library was used for solving ODEs.
%
Parameters $t_\Min$, $\epsilon$, and $\theta$ should be configured.
Each parameter corresponds to the smallest integration step size that CAPD can take, the threshold used in Figure~\ref{a:searchzero}, or a threshold used in $\AlgName{Inflate}$.
In the experiments, these parameters were set as $t_\Min := 10^{-14}$, $\epsilon := 10^{-14}$, and $\theta := 0.01$.
%

For most of models, our implementation only accepts small interval values, as reported in the next section, because an interval enclosure of the state after a continuous evolution and a jump expands by quite a large amount, and thus, the verification process in $\SearchZero$ or the solving process of CAPD will fail.


\subsection{Bouncing ball}

Example~\ref{ex:bb:bltl} can be verified using the implementation by limiting the initial value of $x_2$ to a small interval of width at most $0.01$.
We verified the model with three configurations:
1) with setting a point initial value to $x_2$;
2) with an interval initial value of width 0.01; and 
3) with a point initial value and the property in which the time bound for the $\Always$ operator was set to $[0,100]$.
Each experiment was run 1000 times with the initial values of $x_2$ randomly picked within $[2,7]$.
The results are shown in Table~\ref{t:ex:bb}.
\begin{table}[h]
\begin{center}
	\caption{\label{t:ex:bb} Experimental results (bouncing ball)} 
    \begin{tabular}{l|r|r|r|r|r|r} \hline
		$\phi$ & width & \# $\Valid$ & \# $\Unsat$ & \# $\Unknown$ & \# errors & time \\
		\hline
		$\Always_{[0,10]} \Eventually_{[0,5]} 2\!-\!x_1 \!<\! 0$
		& 0 & 330 & 670 & 0 & 0 & 0.1s \\
		$\Always_{[0,10]} \Eventually_{[0,5]} 2\!-\!x_1 \!<\! 0$
		& 0.01 & 87 & 10 & 903 & 903 & 0.1s \\
		$\Always_{[0,100]} \Eventually_{[0,5]} 2\!-\!x_1 \!<\! 0$
		& 0 & 186 & 20 & 794 & 794 & 0.5s \\
		\hline
	\end{tabular}
\end{center}
\end{table}
In the table, the second column represents the width of the initial values. For each $r \in \{\Valid, \Unsat, \Unknown\}$, the column ``\# $r$'' represents the number of runs that resulted in $r$.
The column ``\# errors'' represents that how many of $\Unknown$ results are caused by an error in the $\SearchZero$ procedure.
The column ``time'' shows average timings.

The computation of each experiment was quite efficient.

Regarding the rate of inconclusive runs in each experiment, all the verifications succeeded in the first experiment.
Because we set the point initial values and the bounded simulation time, considered trajectories were always enclosed with tight intervals, and thus the verification process succeeded even in a situation that was close to singular.
The result also implied that we did not meet a zeno behavior.
Contrastingly, around 90\% and 80\% of the runs in the second and third experiments, respectively, resulted in errors.
Our verification process with the interval Newton failed more often if a trajectory and a guard approached or they became close to tangent.
Because the model was chaotic, a coarser enclosure of states or a longer simulation time increased the possibility to meet such situations.

In the second and third experiments, most of the successful runs resulted in $\Valid$.
These runs became stable (i.e., there are less difference between the continuous trajectories in each step) in the later steps, and thus satisfied the property $\phi$.
We confirmed that most of the inconclusive runs fell into zeno behaviors.

It was quite rare that the result of $\AlgName{AnalIntervals}$ became $\Unknown$ because there were always a few (or no) tight boundaries in $\T_\phi$ and they rarely contained zero.

\begin{figure}[t]
  \centering
  \subfigure[Considered model] { \includegraphics[height=0.15\textheight]{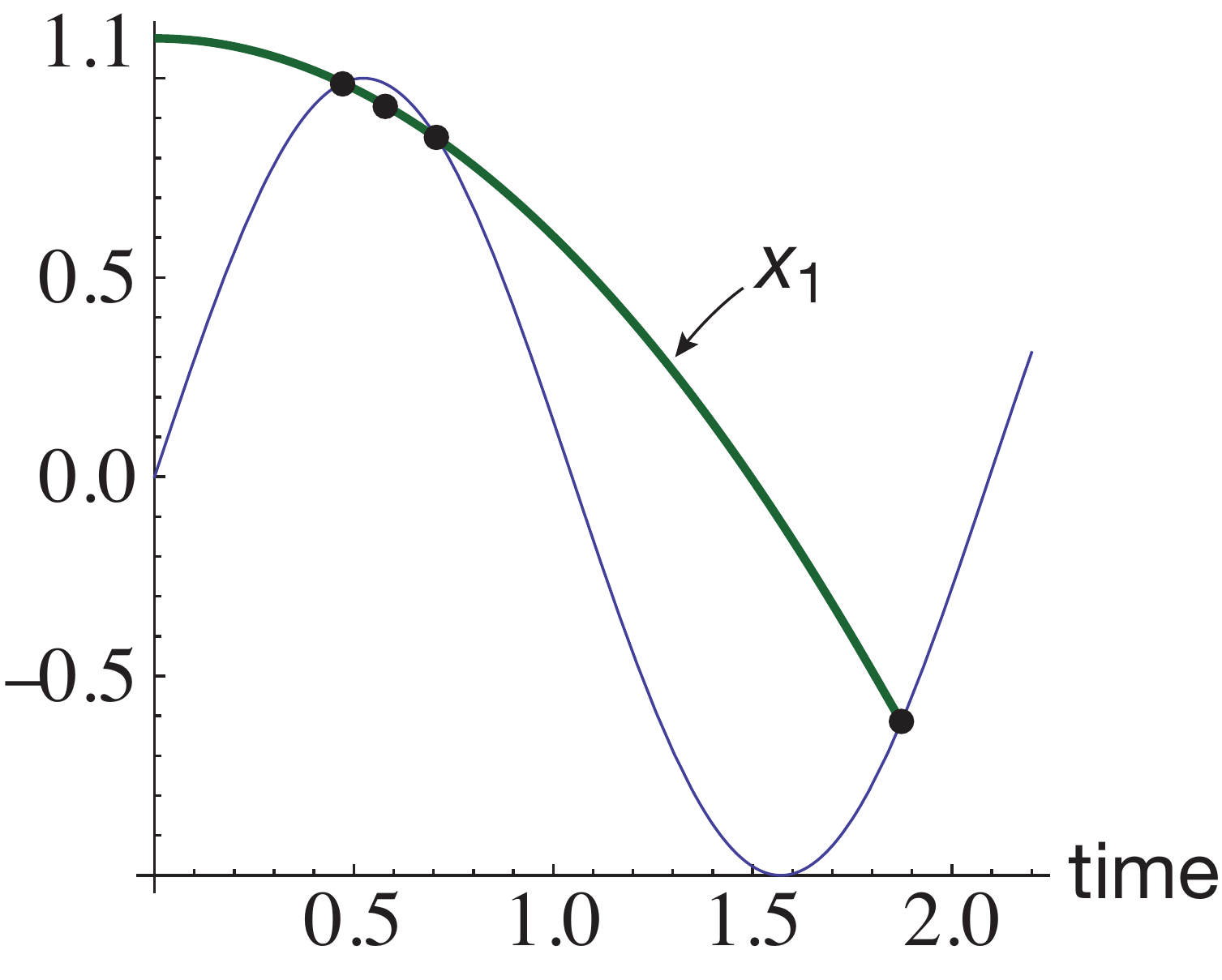}}
  \hspace{3em}
  \subfigure[Our method] { \includegraphics[height=0.15\textheight]{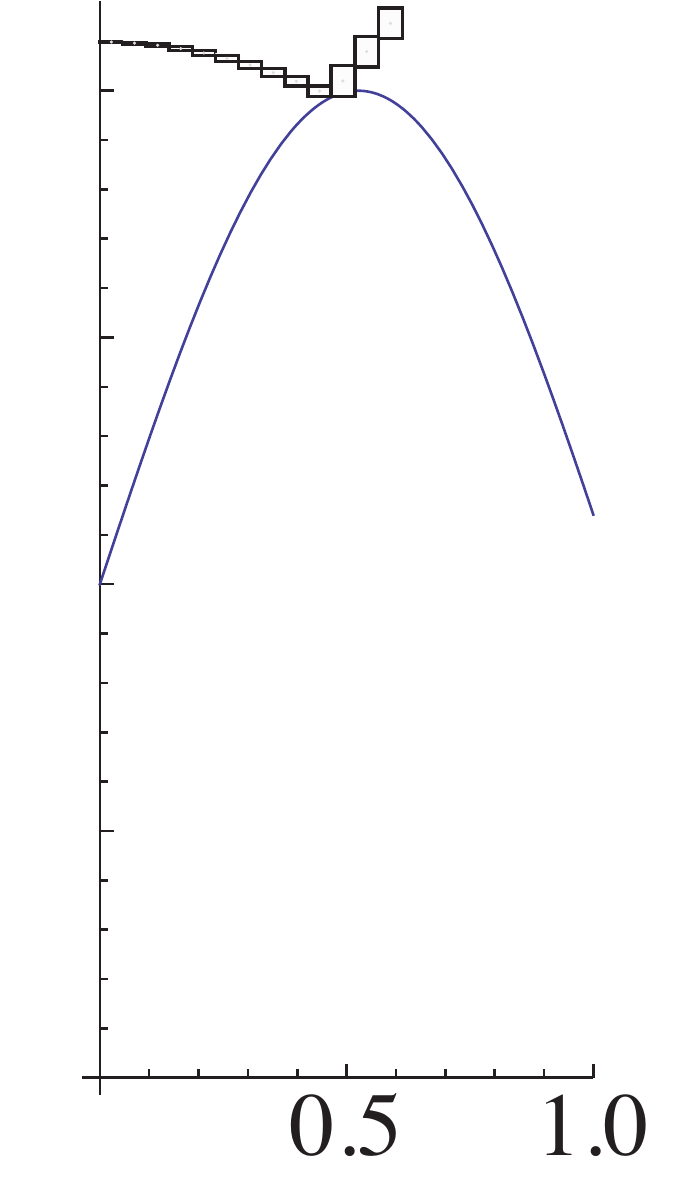}}
  \hspace{3em}
  \subfigure[dReal] { \includegraphics[height=0.15\textheight]{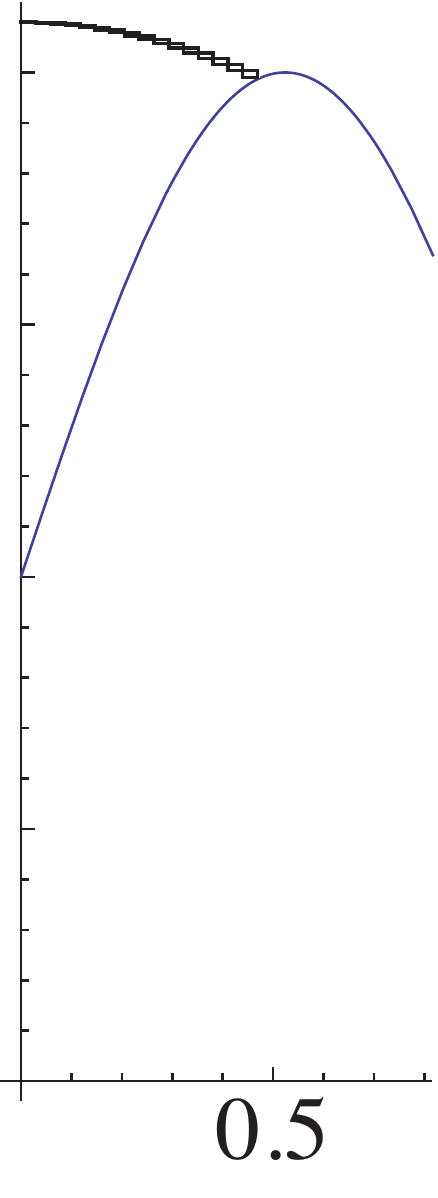}
  \includegraphics[height=0.15\textheight]{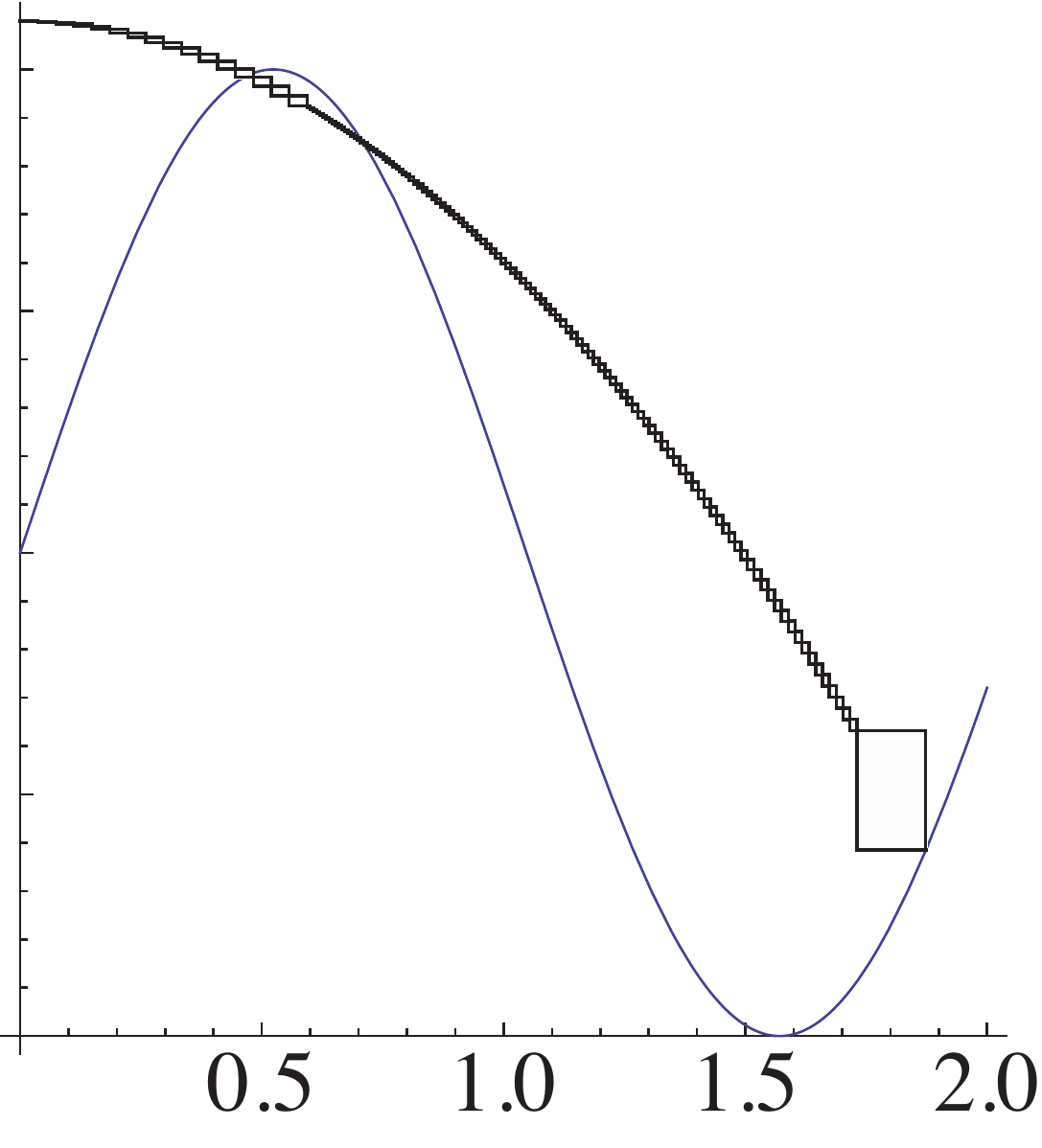} }
\caption{Results of boundary detection of the bouncing ball example}
\label{f:ex:bb}
\end{figure}

Next, we experimented with dReal (version~2.14.08), a solver for $\delta$-weakened SMT problems, for comparison.
We consider a portion of the model of (another instance of) the bouncing ball such that the initial state is $(1.1,0,0)$ and the trajectories of the ball and the table become close to tangent (Figure~\ref{f:ex:bb} (a)).
Next, we analyzed this model by simulating the underlying $\HA$ with our method and by solving the SMT problem with dReal, respectively.
Figure~\ref{f:ex:bb} (b) and (c) show the computed witness trajectories.
Our implementation verified the occurrence of the first contact with the floor.
dReal computed two enclosures for the possible trajectories of the model; they seemed corresponding to the first and last intersections with the guard.
The second witness seemed wrong because the guard condition became $\delta$-$\Sat$ around $t'=0.5$ with $\delta=0.001$.
The computation of a boundary can be troublesome in this way, without the unique existence verification process.

\subsection{Air Traffic Maneuver}



\begin{figure}[t]
  \centering
  \begin{minipage}{.2\textwidth}
    \centering
    \includegraphics[height=0.23\textheight]{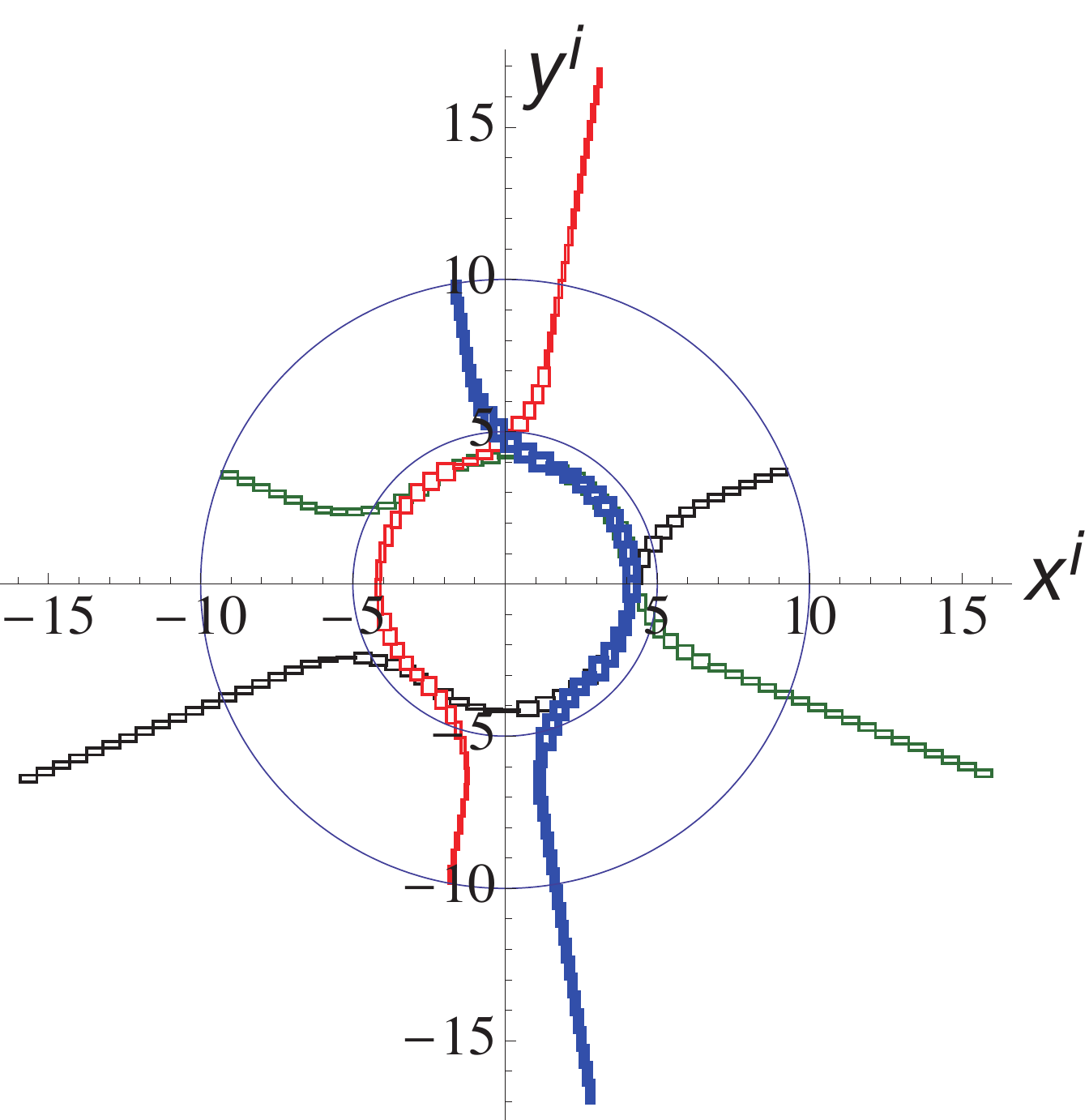}
  \end{minipage}
  \hspace{7em}
  \begin{minipage}{.5\textwidth}
    \centering
	\small
	\begin{align*}
		& \bigwedge_{i=1}^{m-1} \bigwedge_{j=i+1}^m \Always_{[0,30]}( -(x^i-x^j)^2-(y^i-y^j)^2 + \frac{64}{m^2} < 0 ) \\
		\land & \bigwedge_{i=1}^m \Eventually_{[0,10]} \bigl( \Always_{[0,10]}( (x^i-c_x)^2+(y^i-c_y)^2 - 25 < 0 ) \\
	   & \hspace{3em} \land \Eventually_{[10,20]}(  -(x^i-c_x)^2-(y^i-c_y)^2 +100 < 0) \bigr)
	\end{align*}
  \end{minipage}
  \caption{A trajectory of ATM (left) and the BLTL property to verify (right)}
  \label{f:ex:atm}
\end{figure}

We performed a verification of a simplified model of an air traffic maneuver (ATM)~\cite{Platzer2009} in which the number of aircrafts was parameterized.
A trajectory of $(x^i,y^i)$ when $m=4$ is illustrated in Figure~\ref{f:ex:atm}.
We verified the following property shown in Figure~\ref{f:ex:atm}.
The first line describes that the distance between each pair of aircrafts is larger than the threshold $8/m$ during $\Norm{\phi}=30$ time units. 
The following of the property describes that all aircrafts reach within the circle with radius 5 within the time interval $[0,10]$, stay there at least 10 time units, and reach outside the circle with radius 10 after another 10 time units.
We verified 10 times for each instance with $m=2,4,6,8$. In each verification, we randomly picked a point initial value.
All runs resulted in $\Valid$.
The specification of the instances and the results are shown in Table~\ref{t:ex:atm}.
\begin{table}[h]
\begin{center}
	\caption{\label{t:ex:atm} Experimental results (ATM)} 
    \begin{tabular}{r|r|r|r||r|r|r|r} \hline
		$m$ & \# vars & \# APs & time &
		$m$ & \# vars & \# APs & time \\
		\hline
		2 & 10 & 5 & 0.5s &
		6 & 26 & 27 & 28s \\
		4 & 18 & 14 & 5.5s &
		8 & 34 & 44 & 98s \\
		\hline
	\end{tabular}
\end{center}
\end{table}
%
The columns ``\# vars'' and ``\# APs'' represent the numbers of variables in $\HA$ and APs in the property.
The average timings rose exponentially as $m$ increased.

\section{Conclusions}

We have presented a sound BLTL validation method that assures that all initialized trajectories satisfy the property.
The proposed method is able to detect a witness trajectory that is verified its unique existence with an interval-based ODE integration and an interval Newton method.
We consider the experimental results are promising for the practical use.

In future work, our method and implementation should be improved to allow large and uncertain initial values.
Examples in a realistic setting should be demonstrated with the implementation.

\paragraph{Acknowledgments.}
\noindent
This work was partially funded by JSPS (KAKENHI 25880008).

\bibliographystyle{entcs}
\bibliography{/Users/ishii/Documents/bib/mitl.bib}

%
%
%
%
%
%
%

\end{document}